\documentclass[prl, aps, amsfonts, twoside, twocolumn, amssymb, superscriptaddress, sort&compress, reprint]{revtex4-1}

\usepackage[latin1]{inputenc}
\usepackage[english]{babel}
\usepackage{lmodern, graphicx, siunitx, textcomp, xcolor, mathtools, nicefrac, overpic}

\setcitestyle{super}

\newcommand{\ket}[1]{\lvert#1\rangle}
\newcommand{\eqn}{}
\newcommand{\fig}{Fig.}
\newcommand{\figs}{Figs.}
\newcommand{\ops}[1]{\hat{#1}}
\newcommand{\pur}{\ensuremath{\wp}}
\newcommand{\sqz}{\ensuremath{\varsigma}}
\newcommand{\sens}{\ensuremath{\sigma}}
\DeclareMathOperator{\var}{Var}

\DeclareMathOperator{\erf}{erf}
\DeclareMathOperator{\e}{e}

\DeclarePairedDelimiter\expct{\langle}{\rangle}
\DeclarePairedDelimiter\robra{\lparen}{\rparen}
\DeclarePairedDelimiter\abs{\lvert}{\rvert}

\DeclarePairedDelimiterX\ketbra[2]{\lvert}{\rvert}{#1 \delimsize\rangle \delimsize\langle #2}

\begin{document}

\title{Deterministic phase measurements exhibiting super-sensitivity and super-resolution}

\author{Clemens Sch\"afermeier}
\email{clemens@fh-muenster.de}
\affiliation{Technical University of Denmark, Department of Physics, Fysikvej 309, 2800 Kongens Lyngby, Denmark}
\affiliation{Kavli Institute of Nanoscience, Delft University of Technology, 2628 CJ Delft, The Netherlands}
\author{Miroslav Je\v zek}
\affiliation{Department of Optics, Faculty of Science, Palacky University, 17.\ listopadu 1192/12, 77146 Olomouc, Czech Republic}
\author{Lars S.\ Madsen}
\altaffiliation[Current address: ]{Centre for Engineered Quantum Systems, School of Mathematics and Physics, University of Queensland, St.\ Lucia, Queensland 4072, Australia}
\author{Tobias Gehring}
\author{Ulrik L.\ Andersen}
\email{ulrik.andersen@fysik.dtu.dk}
\affiliation{Technical University of Denmark, Department of Physics, Fysikvej 309, 2800 Kongens Lyngby, Denmark}

\begin{abstract}
  Phase super-sensitivity is obtained when the sensitivity in a phase measurement goes beyond the quantum shot noise limit, whereas super-resolution is obtained when the interference fringes in an interferometer are narrower than half the input wavelength.
  Here we show experimentally that these two features can be simultaneously achieved using a relatively simple setup based on Gaussian states and homodyne measurement.
  Using \num{430} photons shared between a coherent- and a squeezed vacuum state, we demonstrate a \num{22}-fold improvement in the phase resolution while we observe a \num{1.7}-fold improvement in the sensitivity.
  In contrast to previous demonstrations of super-resolution and super-sensitivity, this approach is fully deterministic.
  
  \textbf{\textcopyright\ \href{https://doi.org/10.1364/OPTICA.5.000060}{Optica Vol.\ 5, Issue 1 (2018), pp.\ 60--64, Optical Society of America.}
   One print or electronic copy may be made for personal use only.
   Systematic reproduction and distribution, duplication of any material in this paper for a fee or for commercial purposes, or modifications of the content of this paper are prohibited.}
\end{abstract}

\maketitle

\section{Introduction}

Quantum interference of light plays a pivotal role in high-precision quantum sensing \cite{Giovannetti2004}, optical quantum computation \cite{Kok2007} and quantum state tomography \cite{Lvovsky2009}.
It is typically understood as two-beam interference which can be observed, for instance, in a Mach--Zehnder interferometer or a double slit experiment.
At the output, such interferometers create an oscillatory pattern with a periodicity given by half of the wavelength ($\nicefrac \lambda 2$) of the radiation field, which may be referred to, in analogy to the resolution-benchmark in optical imaging, as the ``Rayleigh criterion'' for phase measurements.
This limit can however be surpassed using different types of states or measurement schemes \cite{Rarity1990, Boto2000, Mitchell2004, Resch2007, Distante2013, Gao2010, Cohen2014}.
In particular, measurement schemes which are based on parity detection \cite{Gerry2000, Gao2010} or approximate parity detection via a phase-space relation \cite{Distante2013} are utilised to beat this limit with classical states, i.e.\ they do not require quantum states
\footnote{A simple approach for increasing the fringe resolution by classical means is a multi-pass interferometer.
As the mentioned techniques can be mapped to multi-pass configurations, we focus on the single-pass configuration.}.
The arguably best-known quantum approach to observe a fringe narrowing uses NOON states, $\ket \psi \propto \ket{N, 0} + \e^{i N \phi} \ket{0, N}$.
Surpassing the Rayleigh criterion is referred to as super-resolution \cite{Jacobson1995, Fonseca1999} and is studied in the context of, e.g., optical lithography \cite{Boto2000}, matter-wave interferometry \cite{Dowling1998} and radar ranging \cite{Jiang2013}.

In quantifying the performance for applications in quantum sensing and imaging, it is common to evaluate the Fisher information \cite{Braunstein1994} or, equivalently, determine the sensitivity in the interferometric phase measurement.
Using coherent states of light the optimal sensitivity is given by $\nicefrac{1}{\sqrt N}$ where $N$ is the mean number of photons of the state \cite{Monras2006}.
This sensitivity constitutes the shot noise limit (SNL).
Overcoming the SNL is commonly referred to as super-sensitivity and can be achieved by non-classical states \cite{Nagata2007, Caves1981, Giovannetti2004, Matthews2016}.
Super-sensitivity based on squeezed states of light has proven to be a powerful and practical way to enhance the sensitivity of gravitational wave detectors \cite{Grote2013, LIGO2013}.

The effects of super-sensitivity and super-resolution can be obtained simultaneously.
For example, optical NOON states offer a sensitivity with Heisenberg scaling, $\nicefrac 1 N$, and a phase resolution that scales as $\nicefrac{\lambda}{2 N}$ corresponding to $N$ fringes per half-wavelength.
NOON states thus exhibit an equal scaling in the two effects.
In contrast, this work shows how resolution and sensitivity are tuneable and can, in fact, compete with each other.
Due to the high fragility of NOON states, the complexity in their generation and the commonly probabilistic way of generation, super-sensitivity and super-resolution have been only measured in the coincidence basis and in a highly probabilistic setting \cite{Mitchell2004, Bouwmeester2004, Nagata2007, Israel2014}.
It has also been suggested to use two-mode squeezed vacuum states in combination with parity detection to attain the two ``super-features'' simultaneously \cite{Anisimov2010}.
However, possibly due to the complications in implementing a parity detection scheme, it has so-far never been achieved experimentally.
The complexity associated with the two schemes are due to the involved non-Gaussian states (NOON states) or the non-Gaussian measurements (parity detection).
A natural question to ask is whether the same ``super-features'' can be realised using simple Gaussian operations.
Here we answer this question in the affirmative.

We propose and experimentally demonstrate that, by using Gaussian states of light and Gaussian measurements, it is possible to realise a phase measurement which features super-resolution and super-sensitivity simultaneously.
Using displaced squeezed states of light in conjunction with homodyne detection followed by a data-windowing technique, we show that the interferometric fringes can be made arbitrarily narrow while at the same time beating the shot noise limit.
In stark contrast to the NOON state scheme which, in any practical setting, is highly probabilistic both in preparation and in detection, our approach provides a deterministic demonstration of super-resolution and super-sensitivity.

\section{Materials and methods}

\begin{figure}
  \centering
  \includegraphics{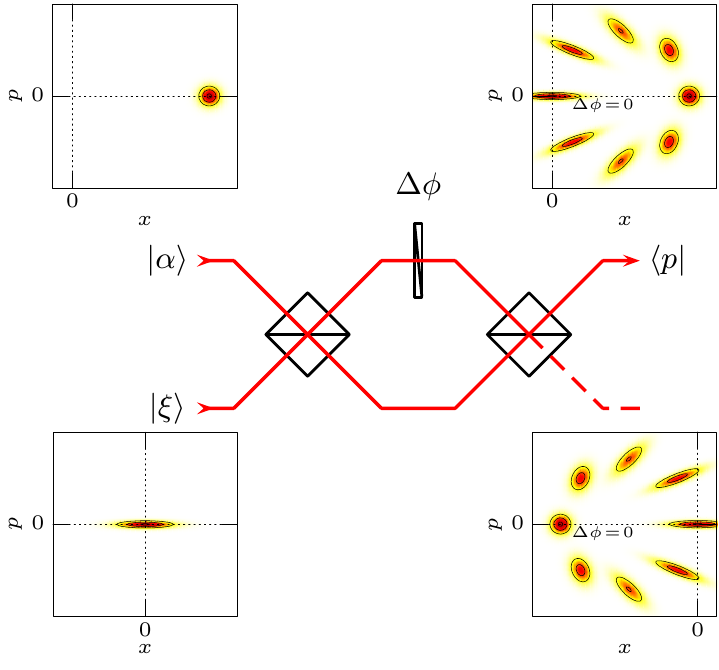}
  \caption{%
  Scheme of the approach.
  A coherent state $\lvert \alpha \rangle$ and a vacuum squeezed state $\lvert \xi \rangle$ are interfered on the first beam splitter.
  Insets show Wigner functions of the respective states, simulated for $\sqz = \nicefrac{1}{\e}$ and $\alpha = 10$.
  In one of the resulting modes, a variable phase shifter is placed.
  At the second beam splitter the modes interfere again, producing the depicted Wigner functions.
  Eight superimposed distributions illustrate the effect of the phase shift, that is each distribution is separated by $\nicefrac \pi 4$.
  If at $\Delta \phi = 0$ the squeezed vacuum state leaves the upper arm, the coherent state exits the lower one.
  Finally, the state is projected onto the quadrature eigenstate $\langle p \rvert$ and partitioned by $\ops \Pi$.
  }%
  \label{fig:scheme}%
\end{figure}

An illustration of the basic scheme is shown in \fig\ \ref{fig:scheme}.
A vacuum squeezed state is combined with a coherent state of light at the entrance to a symmetric Mach--Zehnder interferometer.
The Wigner function at the input is given by
\begin{multline}
  W_\text{in}(x_1, p_1, x_2, p_2) =
  W_{\ket \alpha}(x_1, p_1) W_{\ket \xi}(x_2, p_2) = \\
  \frac{2 \exp\robra*{-2 \robra*{(x_1 - \alpha)^2 + p_1^2}}}{\pi} \cdot
  \frac{2 \pur \exp\robra*{-2 \robra*{\pur^2 \sqz^2 x_2^2 + \frac{p_2^2}{\sqz^2}}}}{\pi},
\end{multline}
where $x_n$ and $p_n$ are the amplitude- and phase-quadratures, $\alpha$ is the amplitude of the coherent state, $\sqz = \e^{-r}$, where $r$ denotes the squeezing parameter, and $\pur$ represents the purity of the squeezed state.
In the scheme, an amplitude modulated coherent state and a phase-squeezed vacuum state interfere on the first beam splitter of the interferometer.
Then the resulting state acquires a relative phase shift $\Delta\phi$, next interferes on the second beam splitter and finally one of the outputs is measured.
As we used weak input signals, a homodyne readout scheme was employed.
\fig\ \ref{fig:scheme} illustrates the trajectory of the output state in phase space for different phase shifts.

If the interferometer is operated near a dark fringe, i.e.\ biasing the phase shift such that most of the light exits the second output of the interferometer, the phase-squeezed vacuum state will be detected.
Thereby, the shot noise around the bias is suppressed and the phase sensitivity improved.
The approach of feeding the commonly unused input mode with a vacuum squeezed state is equal to the proposal by Caves \cite{Caves1981} to beat the shot noise limit in phase measurements.
However, since the phase response for Caves' scheme reads $N \cos^2(\nicefrac \phi 2)$, which is an oscillating function with a period equal to $\nicefrac \lambda 2$, the resolution coincides with the mentioned ``Rayleigh criterion'' for phase measurements.
In the following we show that by implementing a homodyne windowing scheme, the setup yields super-resolution and super-sensitivity.

\begin{figure*}
  \begin{overpic}{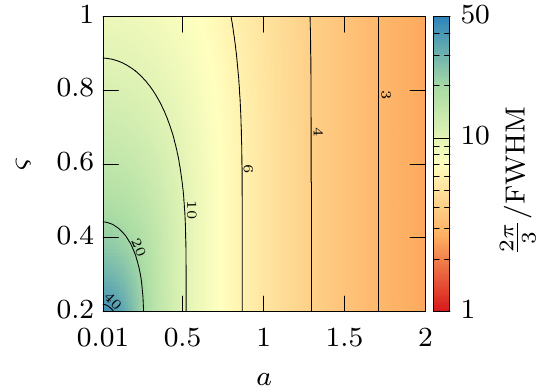}%
  \put(1, 65) {\textbf{a)}}%
  \end{overpic}%
  \hfill\ 
  \begin{overpic}{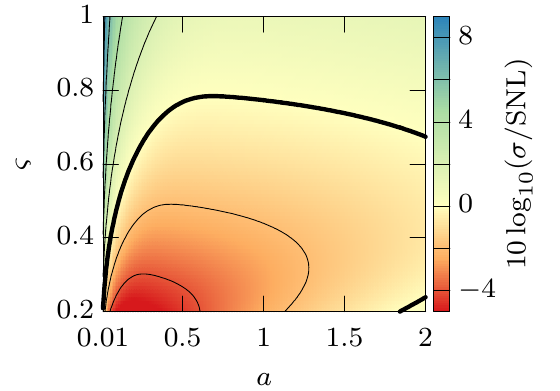}%
  \put(1, 65) {\textbf{b)}}%
  \end{overpic}%
  \hfill\ 
  \begin{overpic}{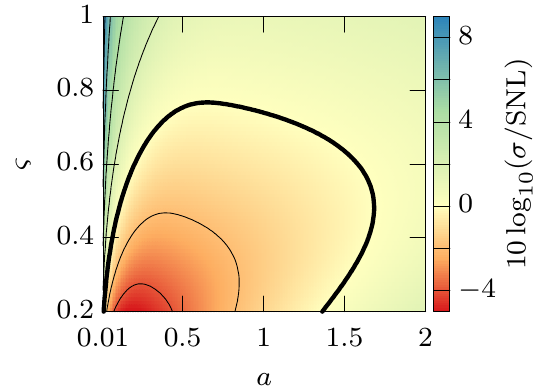}%
  \put(1, 65) {\textbf{c)}}%
  \end{overpic}%
  \caption{%
  Performance of the protocol for $\alpha = 10$ under variation of the bin size and squeezing parameter.
  \textbf{a)} Improvement of the FWHM compared to the Rayleigh criterion ($\nicefrac{2 \pi}{3}$).
  The achieved FWHM is extracted numerically from the response of a pure state.
  The improvement is monotonic in the sense that a smaller bin size $a$ always leads to a higher resolution.
  \textbf{b)} Maximum sensitivity compared to the SNL.
  The region with negative values describes the parameter space where the SNL is surpassed.
  The threshold is marked by the bold line.
  \textbf{c)} Unlike case b), we set the purity $\pur = \nicefrac 1 2$.}
  \label{fig:maps}
\end{figure*}

The quadrature measurement of the homodyne detector is divided into two bins, set by the `bin size' $a$:
If the phase quadrature $\ops p$ is measured, we categorise two different results which are associated with the intervals $\abs p \leq a$ and $\abs p > a$.
We describe such a measurement strategy by the projectors
\begin{equation}
  \ops{\Pi}_0 = \int_{-a}^{a} \text d p \; \ketbra{p}{p},
  \quad
  \ops{\Pi}_1 = \ops{\text I} - \ops{\Pi}_0.
  \label{eq:dichotomy_operator}
\end{equation}
The measurement observable can thus be written as $\ops \Pi = \lambda_0 \ops{\Pi}_0 + \lambda_1 \ops{\Pi}_1$, where $\lambda_0 = 1 / \erf(\sqrt 2 a)$ and $\lambda_1 = 0$ are the eigenvalues associated with the two measurement outcomes.
Now the detector response is found by evaluating $\expct{\ops \Pi}$ which, in the idealised case of $\ops \Pi = \ketbra{p = 0}{p = 0}$, i.e.\ $a \to 0$, and a pure squeezed vacuum state, yields
\begin{multline}
  \expct{\ops \Pi}_{a \to 0, \pur = 1} = \\
  \frac{2 \sqz^2 \exp \robra*{
  -\frac{2 \sqz^2 \abs{\alpha}^2 \sin^2\phi}
  {2 \sqz^2 \robra*{\sqz^2 - 1} \cos \phi + \robra*{\sqz^4 - 1} \cos^2 \phi + \robra*{\sqz^2 + 1}^2}
  }}
  {\sqrt{\robra*{\sqz^4 - 1} \cos^2 \phi + 2 \robra*{\sqz^2 - 1} \sqz^2 \cos \phi + \robra*{\sqz^2 + 1}^2}}.
\end{multline}
The full-width-half-maximum (FWHM) of this function follows $\nicefrac{1}{\abs \alpha}$ for $\abs \alpha \to \infty$, thereby indicating that the interference fringes become narrower as $\alpha$ is increasing and thus demonstrating super-resolution.
It should be stressed that setting $a = 0$ is an idealisation, as it means a projection on an infinitely squeezed state, i.e.\ even number state.
However it points out that the operator $\ops \Pi$ is in some sense an approximation of the parity operator \cite{Gerry2000, Gao2010}.
Considering instead a realistic setting where $a \neq 0$ and the squeezed state is not pure ($\pur < 1$), the response function reads
\begin{equation}
  \expct{\ops \Pi} =
  \frac{1}{2 \erf\robra*{\frac{\sqrt 2 a}{\sqz}}}
  \left[\erf \robra*{\sqrt \frac{2}{c_1} c_-} + \erf \robra*{\sqrt \frac{2}{c_1} c_+}\right],
\end{equation}
where $c_\pm = a \pm \frac 1 2 \abs \alpha \sin \phi$ and
\begin{equation}
  c_1 = \frac{\pur^2 \sqz^2 \robra*{\sqz^2 (\cos \phi + 1)^2 + 2 (1 - \cos\phi)} - \cos^2 \phi + 1}
  {4 \pur^2 \sqz^2}.
\end{equation}
The scaling of the FWHM is preserved for a general value $a$, i.e.\ FWHM $\propto \nicefrac{1}{\abs \alpha}$.
In \fig\ \ref{fig:maps}a we plot the FWHM-improvement as a function of the squeezing parameter $\sqz$ and the bin size $a$.
It is clear from this plot that the super-resolution feature only depends weakly on the degree of squeezing, and a similar conclusion is found for the purity of the state.
The only critical parameter for attaining high resolution is the mean photon number of the input coherent state.
More details and a derivation may be found in the Supplemental Document.

We now turn to the investigation of the sensitivity using the above scheme.
The sensitivity can be found using the uncertainty propagation formula,
\begin{equation}
\sens = \Delta \ops \Pi \big/ \abs[\big]{\nicefrac{\text d}{\text d \phi} \expct{\ops \Pi}},
\label{eq:error_propagation}
\end{equation}
where $\Delta \ops \Pi = \sqrt{\expct{\ops \Pi^2} - \expct{\ops \Pi}^2}$, and for our measurement operator it follows
\begin{multline}
  \label{eq:sensitivity}
  \sens =
  \left| c_1^{\nicefrac 3 2} \sqrt{(2 - c_2) c_2 \pi / 2} \Big/ \right. \\
  \left. \left[\exp\robra*{-2 c_-^2 / c_1} (c_1' c_- + \alpha c_1 \cos \phi) + \right. \right. \\
  \left. \left. \exp\robra*{-2 c_+^2 / c_1} (c_1' c_+ - \alpha c_1 \cos \phi)\right] \right|,
\end{multline}
with the notation $c_1' = \frac{\text d}{\text d \phi} c_1$ and $c_2 = \erf \robra[\big]{\sqrt{\nicefrac{2}{c_1}} c_-} + \erf \robra[\big]{\sqrt{\nicefrac{2}{c_1}} c_+}$.
For a specific parameter regime defined by the purity $\pur$, the bin size $a$ and the squeezing parameter $\sqz$, this sensitivity beats the shot noise limit.
In \figs\ \ref{fig:maps}b,c we plot the sensitivity $\sens$ relative to the shot noise limited sensitivity as a function of the bin size and the squeezing parameter for two different purities.
It is shown in \fig\ \ref{fig:maps}c that it is possible to achieve super-sensitivity in a setting where the squeezed state is impure.
In conclusion, both super-sensitivity and super-resolution can be achieved in a practical setup for the parameter space shown in \fig\ \ref{fig:maps}c.
Furthermore, sensitivity and resolution features are neither independent nor fixed with respect to each other, but can be varied by the homodyne windowing technique.
A discussion of the ultimate sensitivity may be found in the Supplemental Document.

\begin{figure}
  \centering
  \includegraphics{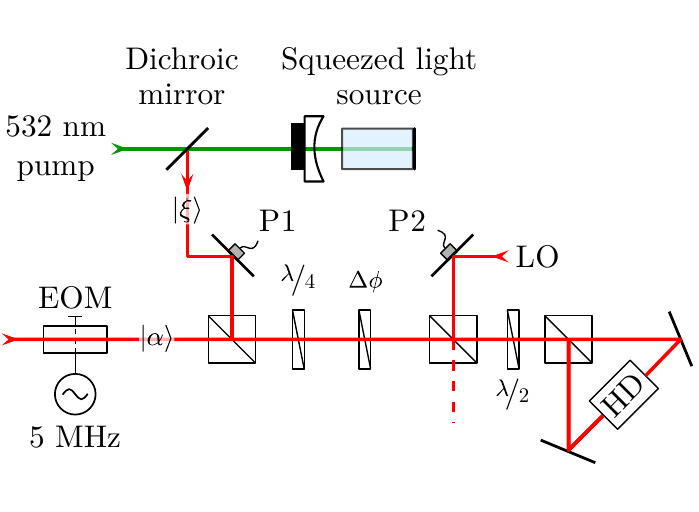}
  \caption{%
  Experimental implementation.
  A vacuum squeezed state, created by parametric down-conversion, and a coherent state, generated via an electro-optic modulator (EOM), are sent into a polarisation-based Mach--Zehnder interferometer (MZI).
  A quarter-wave plate in combination with a motorised half-wave plate ($\Delta\phi$) forms the equivalent phase shift of a MZI where the two modes are spatially separated.
  The piezo transducers P1 and P2 stabilise the phase between the input states and the local oscillator (LO), respectively.
  A half-wave plate in front of the last polarising beam splitter is used to balance the photocurrent in the homodyne detector (HD).
  All cubes represent polarising beam splitters.
  }%
  \label{fig:setup}%
\end{figure}

We proceed by discussing the experimental realisation depicted in \fig\ \ref{fig:setup}.
A squeezed vacuum state and a coherent state with a controllable photon number is injected into the input ports of a polarisation based Mach--Zehnder interferometer.
The polarisation basis ensures high stability and quality of the interference.
Furthermore it allows for simple control of the relative phase shift.
The phase shift is varied by a half-wave plate mounted on a remote-controlled rotation stage.
One output of the interferometer is measured with a high-efficiency homodyne detector exhibiting an overall quantum efficiency of \SI{93}{\percent}, given by \SI{99}{\percent} efficiency of the photo diodes and \SI{97}{\percent} visibility to the local oscillator (LO).
The relative phase of the two input beams of the interferometer as well as the phase of the LO is actively stabilised via real-time feedback circuits, thereby recreating the scheme in \fig\ \ref{fig:scheme} and projecting the output on the $\ops p$ quadrature.
A detailed description may be found in the Supplemental Document.

Squeezed vacuum is generated by parametric downconversion in a \SI{10}{\milli\meter} long periodically-poled KTP crystal embedded in a \SI{23.5}{\milli\meter} long cavity comprising a piezo-actuated curved mirror and a plane mirror integrated with end-facet of the crystal.
A Pound--Drever--Hall (PDH) scheme is adopted to stabilise the cavity resonance.
The downconversion process is pumped by a \SI{45}{\milli\watt} continuous-wave laser beam operating at \SI{532}{\nano\meter}, such that squeezed light is produced at \SI{1064}{\nano\meter}.
To stabilise the pump phase, the radio-frequency signal used also for cavity stabilisation is down-mixed with a phase shift of \SI{90}{\degree}.
Using a \SI{5}{\milli\watt} local oscillator, we observe \SI{6.5 \pm .1}{\decibel} shot noise suppression at \SI{5}{\mega\hertz} sideband frequency, while the anti-squeezed quadrature is \SI{11.3 \pm .1}{\decibel} above shot noise.
The squeezed state parameters read, on average, $\pur = \num{0.58}$ and $\sqz = \num{0.47}$.
A complete characterisation of the squeezed light source is presented in the Supplemental Document.

The coherent input state is produced by an electro-optical modulator (EOM) at a sideband-frequency of \SI{5}{\mega\hertz}.
The chosen frequency ensures the creation of a coherent state far from low-frequency technical noise and with an amplitude $\abs{\alpha}^2$ that is conveniently controlled by the modulation depth of the EOM.

To measure the interferometer's output state at \SI{5}{\mega\hertz}, the electronic output of the homodyne detector is down-mixed at this frequency, subsequently low-pass filtered at \SI{100}{\kilo\hertz} and then digitised with \SI{14}{bit} resolution.
For each phase setting, \num[retain-unity-mantissa = false]{1e6} samples are acquired at a sampling rate of \SI{0.5}{\mega\hertz}.
The data is recorded on a computer for post-processing which includes the dichotomic windowing strategy given by \eqn \eqref{eq:dichotomy_operator} in which we set the bin size $a = \nicefrac 1 2$.
After dividing the data according to $a$, we calculate $\expct{\ops \Pi}$ as well as the standard deviation for each phase setting from the data.
Finally, $\sens$ is computed according to \eqn \eqref{eq:error_propagation}.
The term $\Delta \ops\Pi$ in \eqn \eqref{eq:error_propagation} is extracted directly from the data.
Instead of calculating the derivative of $\expct{\ops \Pi}$ also directly, it is estimated from the theoretical model of $\expct{\ops \Pi}$ fitted to the data.
This approach is chosen to increase the confidence in the computation of $\sens$ and a comparison between this and a direct evaluation is shown in the Supplemental Document.
In the panels on the right of \fig\ \ref{fig:exp_runs}, $\sens$ is shown in comparison to the theoretical model given by \eqn \eqref{eq:sensitivity}.

\section{Results and discussion}

\begin{figure}
  \begin{overpic}{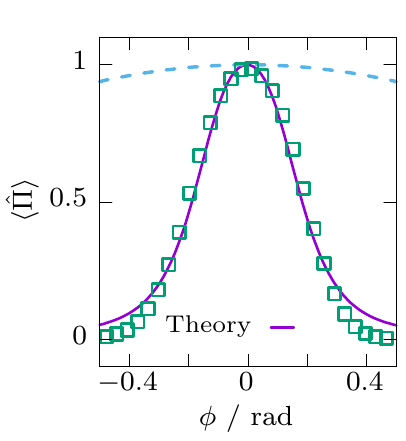}%
  \put(10, 95) {\textbf{a)}}%
  \end{overpic}\ 
  \includegraphics{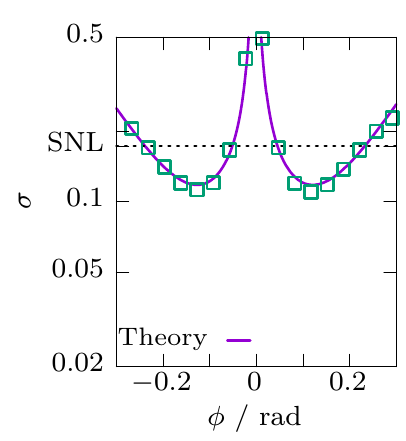} \\
  \begin{overpic}{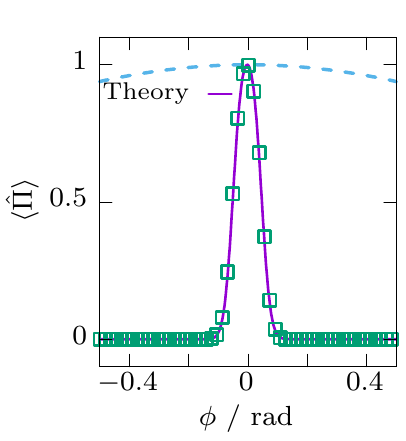}%
  \put(10, 95) {\textbf{b)}}%
  \end{overpic}\ 
  \includegraphics{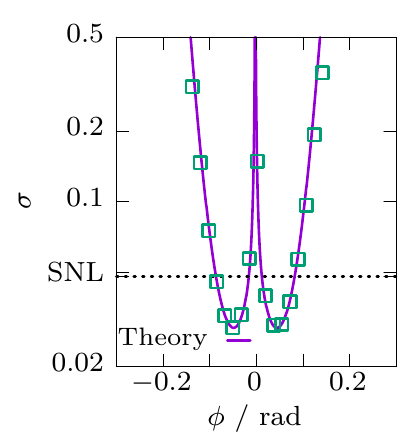}
  \caption{%
  Results achieved for an input state with \textbf{a)} $\abs{\alpha}^2 \approx 30.7$ ($N \approx 33.6$) and \textbf{b)} $\abs{\alpha}^2 \approx 427$ ($N \approx 430$).
  Left: The fringe after applying the dichotomy operator $\ops \Pi$.
  A dashed line follows the fringe of a standard interferometer.
  Its FWHM is \textbf{a)} \num{5.7} and \textbf{b)} \num{22.2} times larger compared to our result.
  Right: The sensitivity derived from the experimental data. In a range of about $\pm \SI{0.1}{\radian}$, the SNL was surpassed by a factor of \textbf{a)} \num{1.5} and \textbf{b)} \num{1.7}.
  The uncertainty of each data point is well within the `$\Box$' symbol.
  We attribute the symmetric deviations at the wings to a systematic anomaly in the set phase-shift controlled by the HWP.
  }%
  \label{fig:exp_runs}%
\end{figure}

\begin{figure}
  \centering
  \begin{overpic}{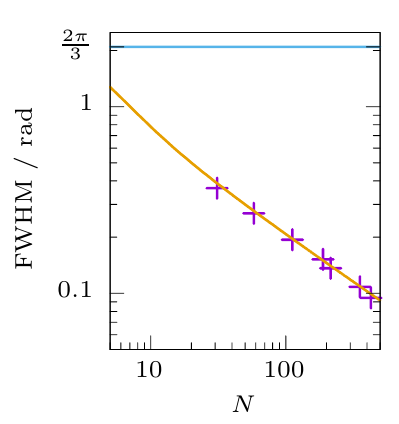}%
  \put(1, 90) {\textbf{a)}}%
  \end{overpic}\ %
  \begin{overpic}{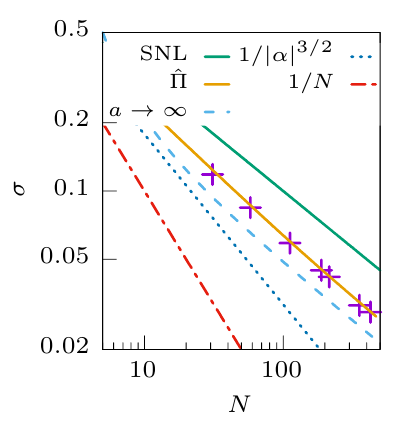}%
  \put(1, 90) {\textbf{b)}}%
  \end{overpic}%
  \caption{%
  Summary of experimental results.
  The solid orange lines are theoretical predictions derived from the measured squeezing parameters and displacement amplitude.
  Each cross symbolises a measurement run.
  The uncertainties are much smaller than the symbol size.
  \textbf{a)} The FWHM under variation of the total average photon number of the input state.
  It always beats the Rayleigh criterion of $\nicefrac{2 \pi}{3}$.
  Comparing the theoretically predicted FWHM proves a stable performance of the setup.
  \textbf{b)} A comparison to four sensitivity limits.
  As for the resolution, the theoretical prediction affirms our experimental results.
  The SNL was outperformed throughout the experiment at a scaling of $N^{-0.56}$; the Heisenberg scaling of $\nicefrac 1 N$ is however not attainable.
  Using no windowing ($a \to \infty$), a better sensitivity and a scaling of $N^{-0.57}$ can be achieved, however it comes at the cost of super-resolution.
  The ultimate bound of our protocol follows $N^{-3/4}$, assuming no losses and restrictions on the squeezing degree.
  }%
  \label{fig:comp}%
\end{figure}

The results for a mean photon number of \num{33.6} and \num{430}, with a mean of \num{2.8} photons contained in the squeezed state and an overall efficiency of circa \SI{84}{\percent}, are shown in \fig\ \ref{fig:exp_runs} and compared with theoretical predictions.
The latter is denoted by solid lines.
It is clear from the plots that the scheme exhibits super-resolution as well as super-sensitivity for certain phase intervals.
We expect that the resolution and sensitivity improves as the mean photon number is increased.
This expectation is confirmed in \fig\ \ref{fig:comp} where the measurement of these two features for increasing photon numbers in the coherent state is depicted.
Specifically, at $\abs{\alpha}^2 = \num{427}$ we obtain a \num{22}-fold improvement in the phase resolution compared to a standard interferometer and a \num{1.7}-fold improvement in the sensitivity relative to the shot noise limit.

It is interesting to compare these results with a scheme exploiting pure optical NOON states which exhibit super-resolution and -sensitivity at the same photon-number scaling.
Using such states, a similar improvement in resolution and sensitivity would require a \num{23}-photon and a \num{3}-photon NOON state, respectively.
Importantly, this only holds for a lossless scenario.
As of today, an optical \num{5}-photon NOON state has been produced which in principle will yield a \num{5}-fold improvement in resolution and a \num{2.2}-fold improvement in sensitivity \cite{Afek2010}.
However, this realisation is intrinsically probabilistic and thus does not exhibit super-sensitivity in a deterministic setting.
To the best of our knowledge, we found that the presented results constitute the first demonstration of super-resolution and super-sensitivity in a deterministic setting.

In summary, we proposed and experimentally demonstrated a simple approach to the simultaneous attainment of phase super-resolution and phase super-sensitivity.
The approach is based on Gaussian squeezed states and Gaussian homodyne measurement followed by a windowing strategy, which is in stark contrast to previously proposed schemes realised with impractical and fragile NOON states, or high-efficiency parity detection.
Our work is of fundamental interest as it highlights the fact that the observation of super-resolution is not a special quantum effect associated with non-Gaussian quantum states \cite{Mitchell2004} or non-Gaussian measurements \cite{Resch2007}.
In conclusion, we find that the actual quantum feature -- that is super-sensitivity -- may co-exist with the super-resolution feature without using advanced non-Gaussian states or non-Gaussian measurements.
Assuming that the measurement's figure of merit is phase sensitivity, we can not find an advantage in exploiting super-resolution in a Gaussian-noise governed context.
Furthermore, we present the trade-off between resolution and sensitivity for the first time and show that significant super-resolution can be achieved at the cost of negligible increase of sensitivity at the scale of a fraction of SNL.
This holds also in the presence of loss and classical Gaussian noise (discussed in the Supplemental Document).
Our result sets a benchmark to evaluate super-resolving strategies, particularly under realistic imperfect conditions.

\section*{Funding Information}
  The work was supported by the Lundbeck Foundation and the Danish Council for Independent Research (Sapere Aude 4184-00338B and 0602-01686B).
  M.J.\ acknowledges support from the Czech Science Agency (project GB14-36681G).

\clearpage

\onecolumngrid

\section{Supplementary Notes}

\subsection{Squeezed state homodyne tomography}

We start by considering the Wigner function of a mixed squeezed vacuum state as
\begin{equation}
  W(x, y) = 2 \pur / \pi \exp\robra*{-2 \robra*{(y / \sqz)^2 + (x \sqz \pur)^2}},
  \label{eq:wigner_mixed}
\end{equation}
where $\sqz \in (0, 1]$ denotes the squeezing parameter and $\pur \in [\sqz, 1]$ the purity.
The squeezing parameter $r$ is related to \sqz\ via $\sqz = \e^{-r}$.

Terming the $y$ axis as the phase quadrature, \eqref{eq:wigner_mixed} describes a phase squeezed state.
Its variances are
\begin{subequations}
  \begin{align}
    \var_x &= \robra*{2 \sqz \pur}^{-2}, \\
    \var_y &= (\sqz / 2)^2,
  \end{align}
\end{subequations}
such that a homodyne tomography parametrised by the phase $\phi$ may be modelled by
\begin{equation}
  \var(\phi, \pur, \sqz) = \frac 1 4 \robra*{\robra*{\frac{\sin(\phi)}{\sqz \pur}}^2 + (\sqz \cos(\phi))^2}.
\end{equation}
In the experiment, we scanned the phase via a piezo-transducer driven by a triangle signal.
To account for the nonlinearity of the piezo-transducer, the model function
\begin{multline}
  u(\phi_1 - \phi)
  \var(a_1 + b_1 \phi + c_1 \phi^2, \pur, \sqz) + \\
  u(\phi - \phi_1) u(\phi_2 - \phi)
  \var(a_2 + b_2 \phi + c_2 \phi^2, \pur, \sqz) + \\
  u(\phi - \phi_2)
  \var(a_3 + b_3 \phi + c_3 \phi^2, \pur, \sqz)
  \label{eq:piezo_model}
\end{multline}
was applied to describe the measured variance.
The coefficients $a, b, c$ provide an approximate description of the piezo's nonlinear behaviour and concatenated step functions $u(\phi)$ recreate the phase intervals $\{\phi_1, \phi_2\}$ caused when the driving signal's slope turns.
From this model function, \sqz\ and \pur\ have been extracted for a given optical pump power.
Before doing so, the recordings were corrected for dark noise and biased with respect to shot noise.
A conversion to a decibel scale was done by $\var^\text{[dB]}_\text{s} = 10 \log_{10}\left(\sqz^2\right)$ and $\var^\text{[dB]}_\text{a} = 10 \log_{10}\left((\sqz \pur)^{-2}\right)$ for squeezing and antisqueezing, respectively.

\subsection{Resolution and sensitivity expression of the protocol}

To arrive at the expression for the response function $\expct{\ops \Pi}$, and in turn the resolution and the sensitivity \sens, we start by propagating the Wigner function (Eq.\ (1) in the main text)
\begin{multline}
  W_\text{in}(x_1, p_1, x_2, p_2, \alpha, \pur, \sqz) =
  W_{\ket \alpha}(x_1, p_1, \alpha) W_{\ket \xi}(x_2, p_2, \pur, \sqz) = \\
  \frac{2 \exp\robra*{-2 \robra*{(x_1 - \alpha)^2 + p_1^2}}}{\pi} \cdot
  \frac{2 \pur \exp\robra*{-2 \robra*{\pur^2 \sqz^2 x_2^2 + \frac{p_2^2}{\sqz^2}}}}{\pi},
\end{multline}
through a Mach--Zehnder interferometer.
According to the main text, $\pur$ and $\sqz$ denote the purity and squeezing parameter, respectively, while the coherent state amplitude $\alpha \in \mathbb R^+$.
The field quadratures are written as $x$ and $p$ and their index represents a certain mode.
The propagation may be described by a combination of a beam splitter, a phase shift and a second beam splitter transformation by
\begin{equation}
  \ops U_\text{I} = \ops U_\text{bs} \ops U_\phi \ops U_\text{bs} =
  \frac 1 2
  \begin{pmatrix}
   \cos \phi + 1 & \sin \phi & \cos \phi - 1 & \sin \phi \\
   -\sin \phi & \cos \phi + 1 & -\sin \phi & \cos \phi - 1 \\
   \cos \phi - 1 & \sin \phi & \cos \phi + 1 & \sin \phi \\
   -\sin \phi & \cos \phi - 1 & -\sin \phi & \cos \phi + 1
  \end{pmatrix}.
\end{equation}
Doing so yields the output modes
\begin{equation}
  \begin{pmatrix}
    x_1' \\
    p_1' \\
    x_2' \\
    p_2'
  \end{pmatrix}
  =
  \frac 1 2
  \begin{pmatrix}
    x_1 - x_2 + (x_1 + x_2) \cos \phi - (p_1 + p_2) \sin \phi \\
    p_1 - p_2 + (p_1 + p_2) \cos \phi + (x_1 + x_2) \sin \phi \\
    -x_1 + x_2 + (x_1 + x_2) \cos \phi - (p_1 + p_2) \sin \phi \\
    -p_1 + p_2 + (p_1 + p_2) \cos \phi + (x_1 + x_2) \sin \phi
  \end{pmatrix},
\end{equation}
where the output quadratures are labelled with a prime.
Next, the Wigner function for the one of the output modes is recovered by tracing out the other one, e.g.\
\begin{multline}
  W(x_1', p_1', \alpha, \pur, \sqz, \phi) = \iint_{-\infty}^\infty \text d x_2' \text d p_2'\ W(x_1', p_1', x_2', p_2', \alpha, \pur, \sqz, \phi) = \\
  4 \pur \sqz \exp\left[
  2 \left(p_1'^2 \robra*{\pur^2 \sqz^2 (\sqz^2 (-\cos^2 \phi) + 2 \cos \phi + \sqz^2 + 2) + c_-^2} +
  2 p_1' \sin \phi \robra*{\alpha c_1 - x_1' c_- (\pur^2 \sqz^4 - 1)} + \right. \right.  \\
  \left. \left. \alpha^2 c_+ c_1 - 2 \alpha x_1' c_+ c_1 + x_1'^2 \robra*{\pur^2 \sqz^2 \robra{\sqz^2 c_-^2 + 2 c_+} - \cos^2 \phi + 1}\right) \Big / \robra*{\robra{\sqz^2 c_- - c_-} c_1} \right] \bigg / \\
  \robra*{\pi \sqrt{\robra*{\sqz^2 (-\cos \phi) + \cos \phi + \sqz^2 + 1} c_1}},
\end{multline}
where $c_\pm = \cos \phi \pm 1$, $c_1 = \pur^2 \sqz^2 c_+ - c_-$.
To arrive at Eq.\ (3) from the main text which represents the idealised case of $\ops \Pi = \ketbra{p = 0}{p = 0}$, i.e.\ $a \to 0$ and $\pur = 1$, one performs the integral
\begin{equation}
  \expct{\ops \Pi}_{a \to 0, \pur = 1} = c_n \int_{-\infty}^\infty \text d x_1' W(x_1', 0, \alpha, 1, \sqz, \phi)
\end{equation}
and norms it with $c_n = \sqz / \sqrt{2 / \pi} $ such that $\expct{\ops \Pi}_{a \to 0, \pur = 1} (\phi = 0) = 1$.
To recover Eq.\ (4) which we used to process the experimental data, one solves the integral
\begin{equation}
  \expct{\ops \Pi} = c_n \int_{-a}^a d p_1' \int_{-\infty}^\infty \text d x_1'\ W(x_1', p_1', \alpha, \pur, \sqz, \phi),
\end{equation}
where $c_n = \robra*{2 \erf\robra*{\sqrt 2 a / \sqz}}^{-1}$.

\subsection{Ultimate sensitivity}

The following arguments can be derived from Eq.\ (7) in the main text.

First, we note that the \emph{resolution} of our protocol improves with increasing the coherent state amplitude $\abs{\alpha}$.
Thus the user aims for the highest laser power which does not corrupt the detection (or the sample).
Then for each coherent state amplitude, the optimum squeezing \sqz, the optimum bin size $a$ and the phase with best sensitivity should be found.
However, the optimum squeezing required is of about \SI{13}{\decibel} already for $\abs{\alpha} = 10$ and increases quickly.
The ultimate sensitivity, given by the maximal Fisher information, then scales as $1 / \abs{\alpha}^{3/2}$, i.e.\ $1 / N^{3/4}$, in the limit of large $N$.

For practical purposes, the optimum can not be reached due to the high degree of squeezing.
The best sensitivity using our resources is reached when $a \to \infty$ which coincides with the limit for squeezed states and standard homodyne detection.

\begin{figure}
  \centering
  \begin{overpic}{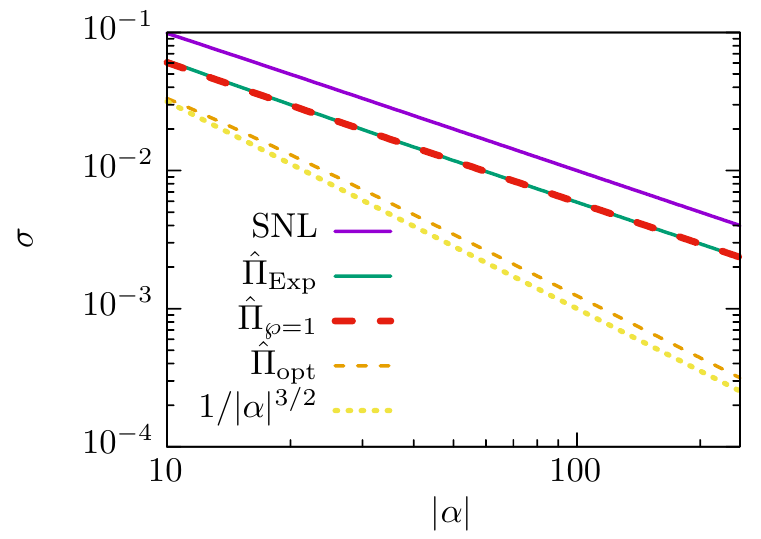}%
  \put(1, 65) {\textbf{a)}}%
  \end{overpic}%
  \hfill\ 
  \begin{overpic}{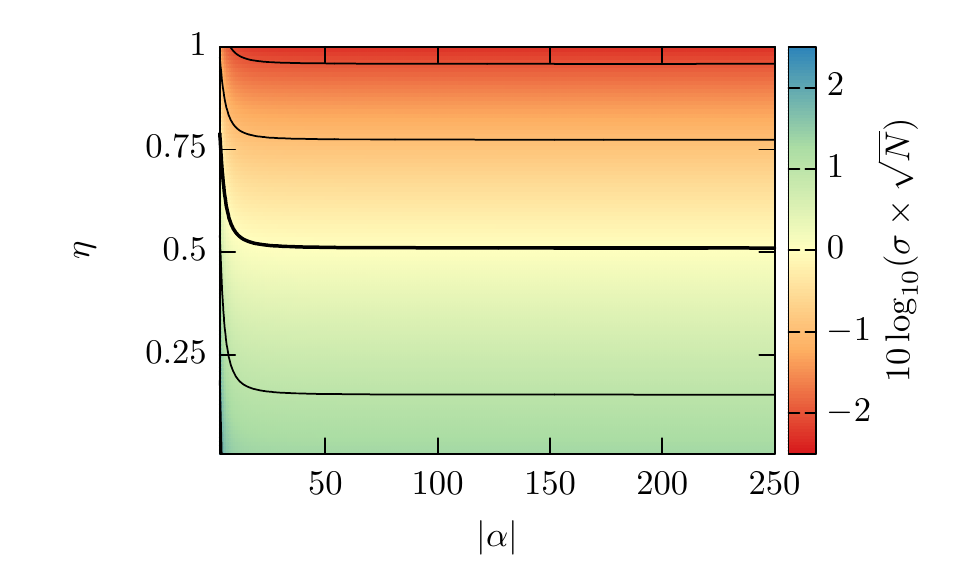}%
  \put(1, 51) {\textbf{b)}}%
  \end{overpic}
  \caption{%
  \textbf{a)} Comparison of $\sens$ for different states under variation of the coherent state amplitude $\abs{\alpha}$.
  The SNL line marks the shot noise limit, given by the number of photons in the coherent state plus the (fixed) number of photons in the squeezed state we used in the experiment.
  Next, $\ops{\Pi}_\text{Exp}$ follows the sensitivity achieved in our experiment.
  To see how much better a pure squeezed state would have performed, $\ops{\Pi}_{\pur = 1}$ was simulated.
  A pure state provides, in our configuration, a negligible enhancement.
  Next, $\ops{\Pi}_\text{opt}$ shows the sensitivity under the assumption that $\pur = 1$ and that the squeezing parameter $\sqz$ is freely tunable.
  The bin size $a$ is fixed to $\nicefrac 1 2$.
  Finally, $1 / \abs{\alpha}^{3/2}$ shows the ultimate sensitivity which is achievable when, in addition to $\pur = 1$ and an optimised $\sqz$, also $a$ is optimised.
  \textbf{b)} A study of the sensitivity improvement when varying the efficiency $\eta$ and $\abs{\alpha}$.
  For this simulation, we assume to use a fixed amount of photons in the squeezed state and a variable amount of photons in the coherent state.
  The ``squeezing'' photons are given by $\sqz$ and $\pur$ in our measurement.
  The total number of photons $N$ is then used to calculate the shot noise limit.
  Negative-valued regions in the contour plot (above the thick contour line) indicate that the shot noise limit is surpassed.
  We find that losses of about \SI{50}{\percent} restrict the measurement from being super-sensitive.
  However, even at lower efficiencies, the feature of super-resolution is still maintained.
  }
  \label{fig:sens_comp}
\end{figure}

In \fig\ \ref{fig:sens_comp}a, various sensitivities are plotted.
The figure shows that the protocol performs quite well with respect to the efficiency, discussed later, of our setup.
Furthermore, it shows that the influence of the bin size $a$, if set to $\nicefrac 1 2$, is small even when the goal is to reach the ultimate bound.

\fig\ \ref{fig:sens_comp}b illustrates the effect of losses for our protocol.
The scenario for this simulation is: given the degree of squeezing ($\sqz \approx 0.47$ or \SI{6.5}{\decibel}) used for the measurements, what is the effect of losses $\eta$ and the coherent state amplitude $\alpha$.
In the Wigner function Eq.\ (1) (main text) we have treated losses in terms of purity $\pur$ which is very convenient when performing state tomography.
For a treatment of losses, it is thus necessary to convert $\pur$ to $\eta$.
This can be achieved by comparing Eq.~(1) to a Wigner function of a squeezed state which suffered losses via a beam splitter transformation.
Once the ancillary modes introduced by the beam splitter are traced out, one may compare the variance of the purity versus the loss based approach and find the mapping
\begin{subequations}
  \begin{align}
    \pur &\to \frac{\sqz}{\sqrt{\left(-\eta \sqz^2 + \eta + \sqz^2\right) \left(\eta \left(\sqz^2 - 1\right) + 1\right)}}, \\
    \sqz &\to \sqrt{\eta \left(\sqz^2 - 1\right) + 1}.
  \end{align}
\end{subequations}
These expressions substitute the original $\sqz$ and $\pur$ parameters in Eq.\ (7).

\subsection{Impact of detector noise on sensitivity and resolution}

So far we have studied the influence of losses and impure squeezing and considered them as effects that occur during the propagation through the interferometer.

\begin{figure}
  \centering
  \begin{overpic}{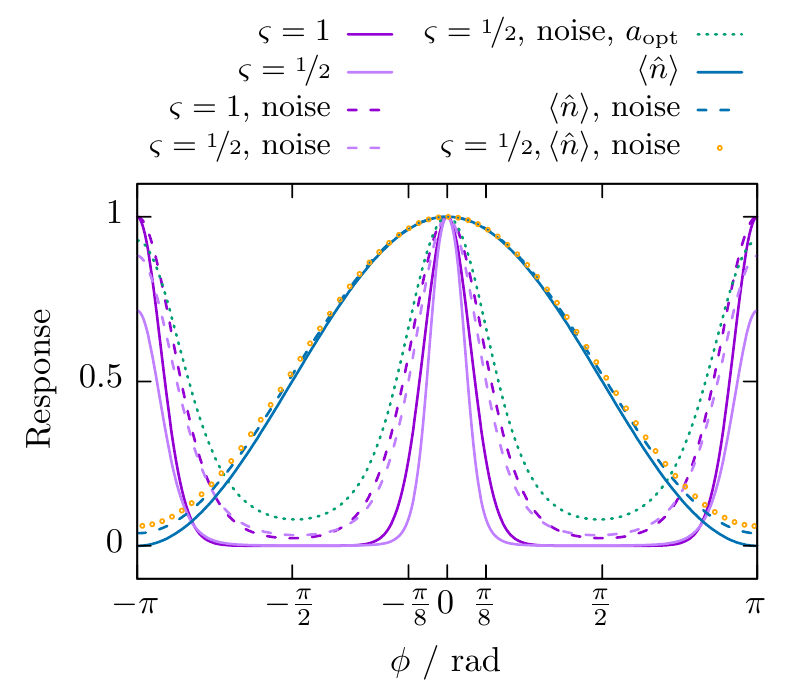}%
  \put(1, 65) {\textbf{a\textsubscript{R})}}%
  \end{overpic}%
  \hfill\ 
  \begin{overpic}{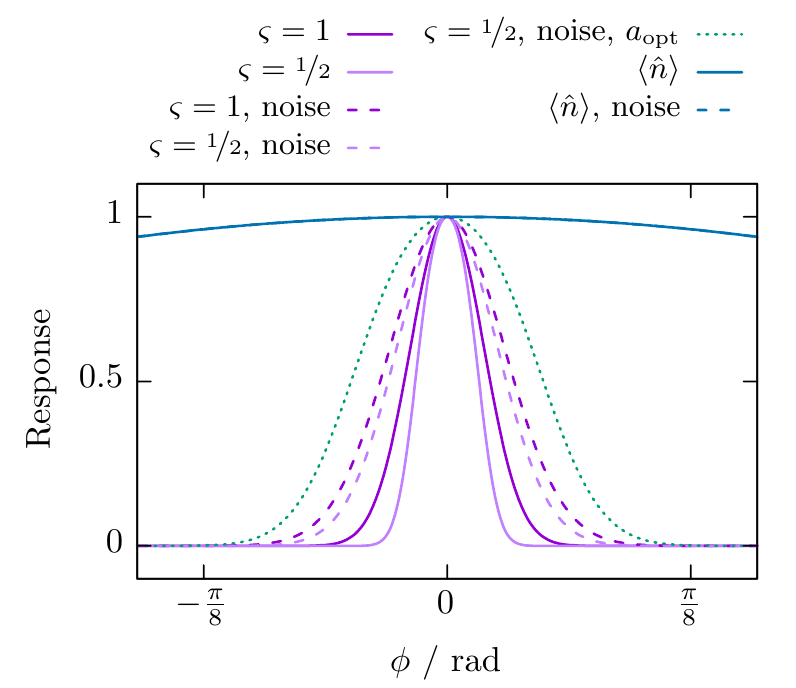}%
  \put(1, 65) {\textbf{b\textsubscript{R})}}%
  \end{overpic} \\
    \begin{overpic}{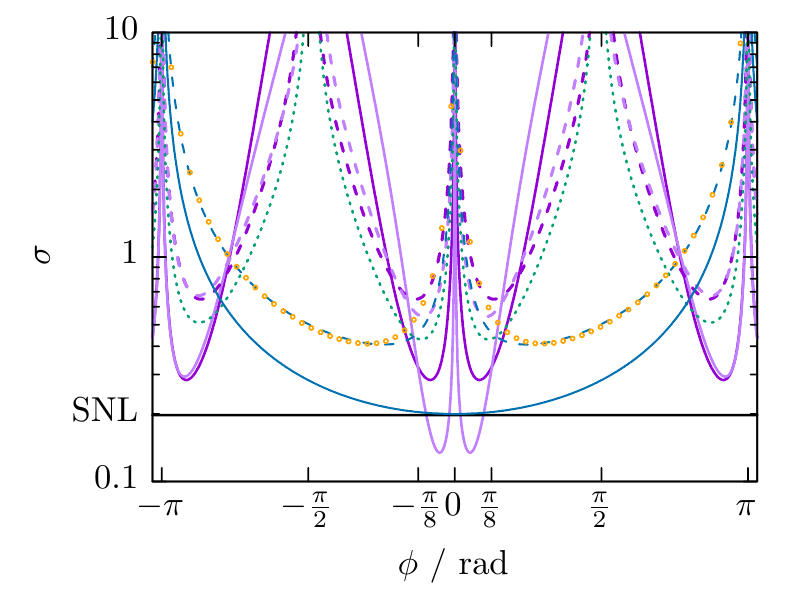}%
  \put(1, 65) {\textbf{a\textsubscript{$\sigma$})}}%
  \end{overpic}%
  \hfill\ 
  \begin{overpic}{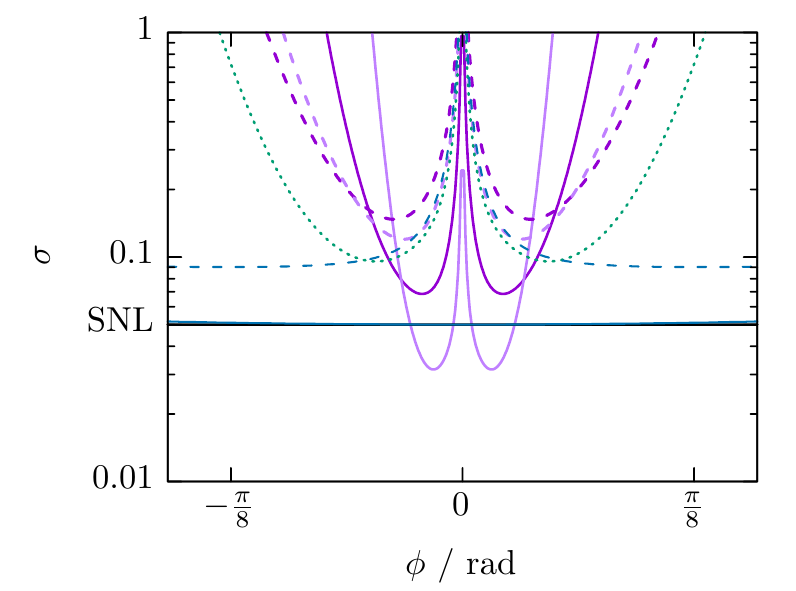}%
  \put(1, 65) {\textbf{b\textsubscript{$\sigma$})}}%
  \end{overpic} \\
  \caption{%
  Interferometer response (top panels) and phase sensitivity (bottom panels) for different detection schemes and input states.
  Comparison for a coherent state amplitude of \textbf{a\_)} $\abs \alpha = 5$ and \textbf{b\_)} $\abs \alpha = 20$.
  Three scenarios are simulated, each with and without added Gaussian detector noise: no squeezing ($\sqz = 1$), squeezing ($\sqz = \nicefrac 1 2$) and direct intensity detection ($\expct{\ops n}$).
  In case of $\sqz = \nicefrac 1 2$, the bin size has been set to $a = \nicefrac 1 2$ and to $a_\text{opt}$ which optimises the sensitivity.
  Furthermore, we assumed a pure squeezed state.
  Panels a\_) also show the scenario of using intensity detection and squeezing, which is a technique suggested by Caves \cite{Caves1981sup}.
  }
  \label{fig:res_w_noise_comp}
\end{figure}

Here we simulate the effect of technical detector noise which enters the mode \emph{after} the output port of the interferometer.
Hence, the thermal photons added by this process do not contribute to the shot noise limit as they do not bear any information about the phase shift.
This study is especially important when the detection is the limiting factor in the system.
To simulate the impact of detector noise, the interferometer's output mode was ``thermalised'' by convoluted it with a Gaussian distribution of unity width.
The results are shown in \fig\ \ref{fig:res_w_noise_comp}.
The ratios of the sensitivity with versus without added noise are summarised in table \ref{tbl:comp}.

First, and similar to the case of losses, our protocol is able maintain super-resolution despite realistic imperfections.
Second, we note that super-resolution does not rely on squeezing, i.e.\ it is a classical effect.
In fact, it can be seen that an increased resolution potentially decreases the sensitivity:
to highlight this finding in this noise context, we simulated the response and sensitivity of our protocol with added noise, first with $a = \nicefrac 1 2$ and second where $a$ was chosen to optimise the sensitivity (denoted $a_\text{opt}$ in the figures).
The case of $a_\text{opt}$ yields a higher sensitivity at the cost of resolution.

\begin{table}%
  \centering
  \begin{tabular}{lccccc}
                     & $\sqz = 1$ & $\sqz = \nicefrac 1 2$ & $\sqz = \nicefrac 1 2, a_\text{opt}$ & $\expct{\ops n}$ & $\sqz = \nicefrac 1 2, \expct{\ops n}$ \\ \hline
  $\abs \alpha = 5$  & 2.3 & 4.0 & 3.2 & 2.0 & 3.3 \\
  $\abs \alpha = 20$ & 2.2 & 3.8 & 3.0 & 1.8 & -- \\ \hline
  \end{tabular}
  \caption{%
  The relative sensitivity-degradation for the schemes shown in \fig\ \ref{fig:res_w_noise_comp}.
  The higher the number, the stronger the impact of detector noise.
  }
  \label{tbl:comp}
\end{table}

In quantum enhanced sensing where sensitivity is the figure of merit and in face of Gaussian noise, we find no advantage in using super-resolution.
The treatment of non-Gaussian detector effects such as thresholds or the effect of analogue-to-digital converters in the signal chain are question left open for future projects.

\section{Supplementary Methods}

\subsection{Basic experimental setup}

\begin{figure}
  \centering
  \includegraphics{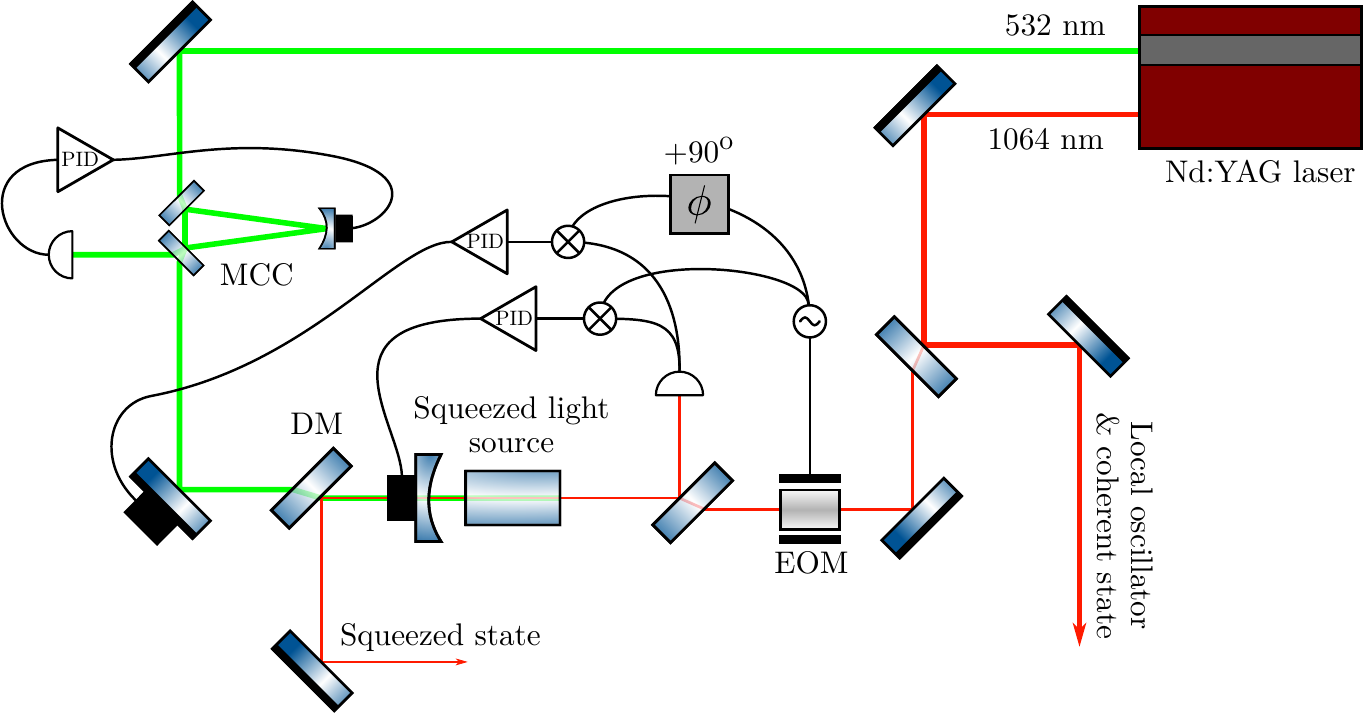}
  \caption{%
  Simplified setup to generate the local oscillator, coherent- and squeezed state.
  MCC: Mode cleaning cavity.
  DM: Dichroic mirror.
  EOM: Electro-optical modulator.
  ESA: Electronic spectrum analyser.
  PID: Servomechanism.
  }
  \label{fig:setup_sup}
\end{figure}

A schematic representation of the basic experimental setup for the generation of squeezed light is shown in \fig\ \ref{fig:setup_sup}.
The setup was powered by a continuous-wave Nd:YAG laser (\textsl{Innolight GmbH} Diabolo) providing \SI{1064}{\nano\meter} and \SI{532}{\nano\meter} radiation, whereas the latter was produced by second harmonic generation from the fundamental.
The second harmonic field was employed as pump for squeezed light generation.
For increased mode matching efficiency, the pump light was filtered by means of a triangular-shaped travelling-wave mode cleaning cavity (MCC) prior to coupling into the squeezing cavity.
The MCC was stabilised using a standard Pound--Drever--Hall (PDH) scheme exploiting the internal phase modulation of the laser at \SI{12}{\mega\hertz} for error signal generation.

The fundamental beam was split into two, one serving as steering beam in the generation of squeezed vacuum state while the second and most intense part was used as local oscillator for homodyne detection and coherent state generation.

In the path of the steering beam, an electro-optic modulator was placed to generate phase modulation sidebands for PDH stabilisation of the squeezer cavity.
A second feedback loop actuating the phase of the pump beam was used for locking the squeezed light source to deamplification or amplification, i.e.\ amplitude- or phase-squeezing, respectively.

\subsection{Characterisation of squeezed light source}

The squeezed-light source consisted of a linear Fabry--P\'erot resonator enclosing a \SI{10}{\milli\meter} periodically poled potassium titanyl phosphate (ppKTP) crystal with two flat end-facets.
One end-facet was high-reflective coated for both wavelengths, the fundamental at \SI{1064}{\nano\meter} and the pump at \SI{532}{\nano\meter}, serving as end mirror for the resonator.
The other end-facet was anti-reflective coated for both wavelengths.
The coupling mirror was attached to a piezo transducer and had a reflectivity of \SI{90}{\percent} for \SI{1064}{\nano\meter} and \SI{20}{\percent} for \SI{532}{\nano\meter}.
The mirror was polished to a radius of curvature of \SI{20}{\milli\meter} and was placed \SI{13}{\milli\meter} from the crystal.
This yielded a full-width-half-maximum cavity bandwidth of about \SI{80}{\mega\hertz}.
To achieve phase matching, the non-linear crystal was attached to a Peltier element.
A phase matching temperature was reached at \SI{36.30}{\celsius}.

The cavity was locked by a PDH phase modulation-demodulation technique at \SI{37.22}{\mega\hertz} using a \SI{550}{\micro\watt} steering beam launched into the cavity from the high-reflective mirror.
The pump phase was locked using the same phase modulation as for the cavity lock, but with a demodulation phase of \SI{90}{\degree} with respect to the cavity lock error signal.
To achieve amplitude squeezing the pump phase was locked to deamplification of the steering beam.

\begin{figure}
  \centering
  \includegraphics{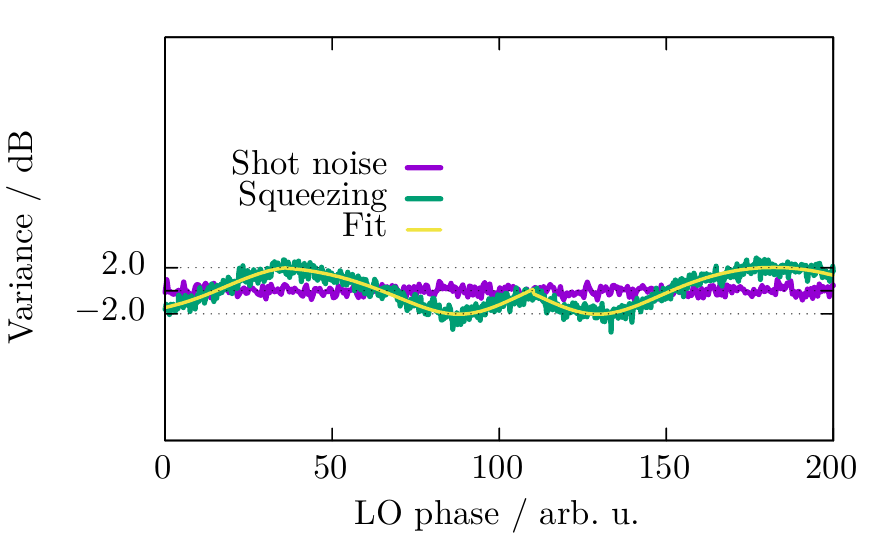} \hfill
  \includegraphics{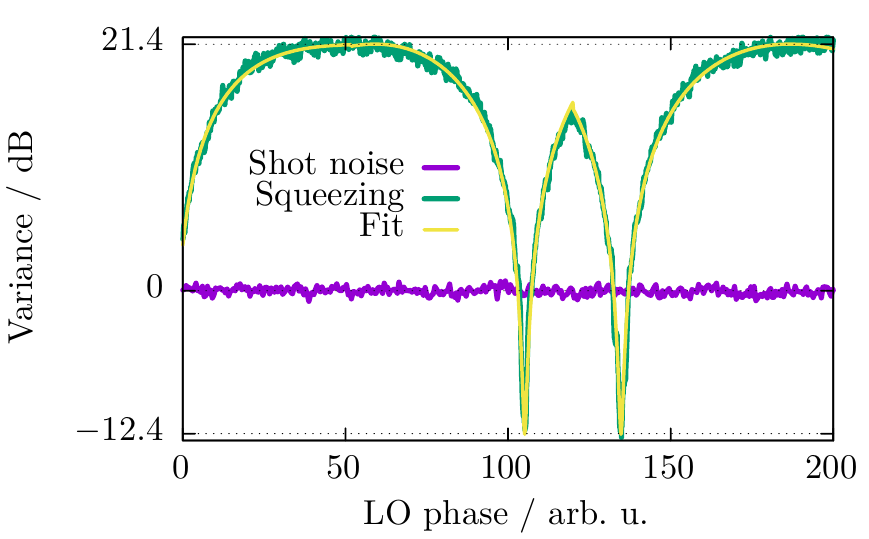}
  \caption{%
  Squeezing spectra at \SI{2}{\milli\watt} (left) and \SI{92}{\milli\watt} optical pump power.
  \num{20000} samples were recorded over \SI{200}{\milli\second} at a resolution bandwidth of \SI{270}{\kilo\hertz} and a video bandwidth of \SI{680}{\hertz} with a \textsl{Agilent} N9000A CXA set to \SI{5}{\mega\hertz}.
  The local oscillator (LO) phase was varied by means of a piezo actuated mirror.
  The light yellow line on top of the experimental data in green was fitted to \eqref{eq:piezo_model}.
  This data was taken downstream the squeezed light source, i.e.\ it bypassed the interferometer used for phase sensing.
  }
  \label{fig:20161129_sqz_figs}
\end{figure}

Prior to sending the squeezed beam into the actual interferometric setup, a characterisation of the squeezing degree was performed.
A homodyne detector was set up close to the output of the squeezed light source to lower optical losses.
For exemplification, \fig\ \ref{fig:20161129_sqz_figs} shows the variance at a pump power of \SI{2}{\milli\watt} and \SI{92}{\milli\watt}.

\begin{figure}
  \centering
  \includegraphics{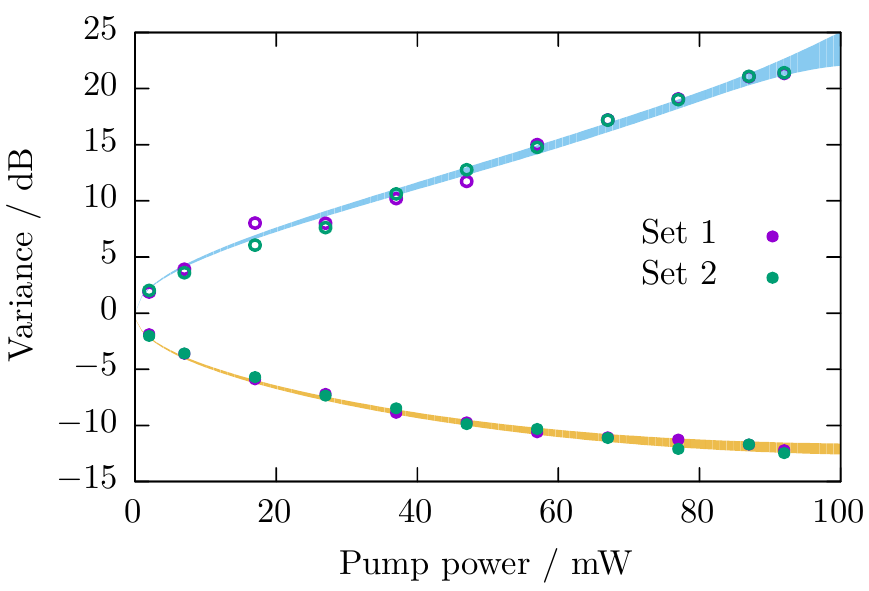}
  \caption{%
  Scaling of (anti-)squeezing variance with respect to pump power.
  Two measurement sets are shown.
  Open circles denote anti-squeezing.
  The filled area represents the lower and upper prediction band of the fit with a confidence level of \SI{99}{\percent}.
  Parameter uncertainties are smaller than the denoting symbols.
  At \SI{92}{\milli\watt} pump power, a shot noise suppression of \SI{12.45 \pm .05}{\decibel} was achieved.
  }
  \label{fig:20161129_pump_vs_var}
\end{figure}

Theoretically, the variance of (anti-)squeezed light versus pump power $p$ follows \cite{Wu1987}
\begin{equation}
  \var_\text{a,s} = \frac 1 2 \robra*{
   \frac{4 \sqrt{\frac{p}{P_\text{th}}} \eta}%
  {4 \robra*{\frac{2 \pi \nu}{\kappa}}^2 + \robra*{1 \mp \sqrt{\frac{p}{P_\text{th}}}}^2} \pm 1},
  \label{eq:sqz_generation}
\end{equation}
where $P_\text{th}$ denotes the threshold power, $\eta$ the total efficiency of the system, $\kappa$ the cavity decay rate and $\nu$ the probing frequency.
The upper sign shall be chosen to describe the antisqueezing variance.
\eqref{eq:sqz_generation} was used to fit the recorded variance of the homodyne tomography and estimate $\eta$, $P_\text{th}$ and $\kappa$.
Inserting the standard deviations of the fit parameters provided the prediction bounds shown in \fig\ \ref{fig:20161129_pump_vs_var}.
Estimated values are summarised in table \ref{tab:srss_fit}.

\begin{table}
  \centering
  \begin{tabular}{l l}
    & Estimate \\ \hline
    $P_\text{th}$ & \SI{113.73 \pm 2.89}{\milli\watt} \\
    $\eta$ & \SI{94.230 \pm 0.251}{\percent} \\
    $\kappa$ & \SI{519.61 \pm 76.87}{\radian \mega \hertz} \\ \hline
  \end{tabular}
  \caption{%
  Estimates with standard uncertainties of the model parameters in \eqref{eq:sqz_generation}.
  Values are based on the two recorded sets shown in \fig\ \ref{fig:20161129_pump_vs_var}.
  }
  \label{tab:srss_fit}
\end{table}

\subsection{Use and efficiency of squeezed light in the interferometer}

In the actual interferometer used for this experiment, the settings in table \ref{tab:srss_settings} apply.
As stated, the pump power was set to \SI{45}{\milli\watt}, i.e.\ only a fraction of the available degree of squeezing was employed.
This choice was made to avoid saturation of the homodyne detector, as to resolve a higher degree of squeezing a stronger local oscillator is required.
When the phase in the interferometer is set such that the squeezed state is detected, this causes no issues.
However as the phase of the interferometer is swept, a much stronger coherent state will be detected.
At high local oscillator power, this easily saturates the electronics.

To characterise the efficiency of squeezed light detection downstream the interferometer, a homodyne tomography similar to the one discussed above was performed.
Theoretically, the loss-reduced degree of squeezing is given by
\begin{equation}
  \var_\text{out}^\text{[dB]} = 10 \log_{10} \robra*{\eta 10^{\var_\text{in}^\text{[dB]} / 10} + (1 - \eta)},
\end{equation}
such that we estimated $\eta \approx \SI{84}{\percent}$ using the values from table \ref{tab:srss_settings}.

\begin{table}
  \centering
  \begin{tabular}{r l p{1em} r l}
    \multicolumn{2}{c}{Set} & & \multicolumn{2}{c}{Measured} \\ \hline
    Parameter & Value & & Parameter & Value \\ \hline
    Crystal temperature & \SI{36.6}{\celsius}    & & Dark noise clearance & \SI{18}{\decibel} \\
    Pump power & \SI{45}{\milli\watt}            & & $\mathcal V(\text{LO} \to \ket \alpha)$ & \SI{99}{\percent} \\
    Steering beam power & \SI{400}{\micro\watt}  & & $\mathcal V(\text{LO} \to \ket \xi)$ & \SI{97}{\percent} \\
    Local oscillator power & \SI{5}{\milli\watt} & & $\var^\text{[dB]}_\text{s}$ & \SI{-6.5 \pm .1}{\decibel} \\
    Down-mix frequency & \SI{5}{\mega\hertz}     & & $\var^\text{[dB]}_\text{a}$ & \SI{11.3 \pm .1}{\decibel} \\
    EOM driving frequency & \SI{5}{\mega\hertz}  & & $\pur$ & \num{.582 \pm .001} \\ \hline
  \end{tabular}
  \caption{%
  Parameter values set or measured during the experiment.
  The visibility $\mathcal V$ was measured for the inference between the local oscillator (LO) and beam for the coherent state ($\ket \alpha$) and squeezed state ($\ket \xi$).
  The degree of (anti-)squeezing and purity $\pur$ were characterised via a fit to a homodyne tomography.
  }
  \label{tab:srss_settings}
\end{table}

\subsection{Interferometer setup}

To perform the quantum enhanced phase measurement scheme presented in the main text, a Mach--Zehnder interferometer was built.
Instead of using the standard configuration, we implemented a polarisation based Mach--Zehnder interferometer where the two spatial modes are substituted by orthogonal polarisation modes.
This means that only one spatial mode comprises the to-be-interfered beams and inherently increases the mechanical stability, as demonstrated experimentally in similar setups \cite{Micuda2014}.

\begin{figure}
  \centering
  \includegraphics{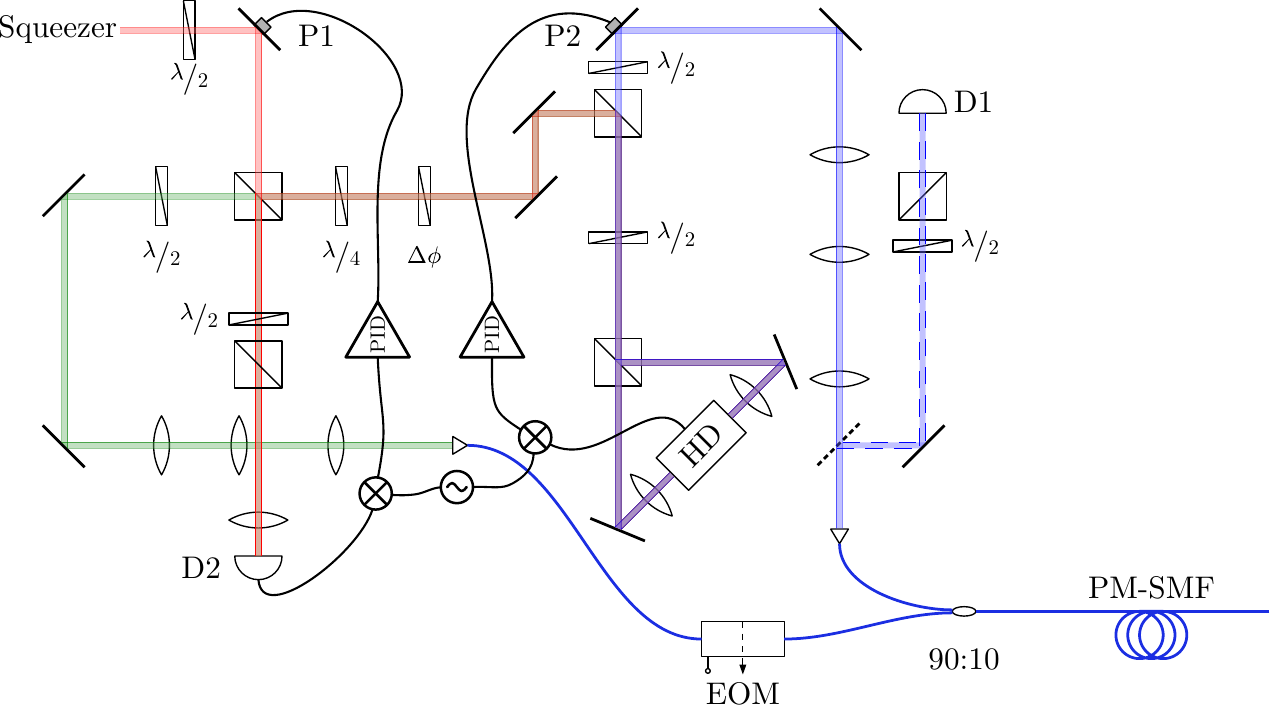}%
  \caption{%
  To-scale schematic of the interferometer.
  All cubes represent polarising beam splitters.
  The wave plate labelled by $\Delta\phi$ was mounted in a remote-controlled rotation stage.
  Detector D1 monitored the power and the polarisation of the local oscillator.
  The interference between a fraction of the coherent- and the squeezed beam was detected at D2.
  Its signal was part of a feedback loop to establish a stable phase stabilisation between the the squeezed- and the coherent state.
  PM-SMF: Polarisation maintaining single-mode fibre.
  HD: Homodyne detector.
  P$n$: Mirror mounted on piezo-transducer.
  EOM: Electro-optical modulator.
  D$n$: Photodetector.
  Red path: Squeezed state.
  Green path: Coherent state.
  Blue path: Local oscillator.
  }
  \label{fig:srss_setup}
\end{figure}

\begin{figure}
  \centering
  \includegraphics{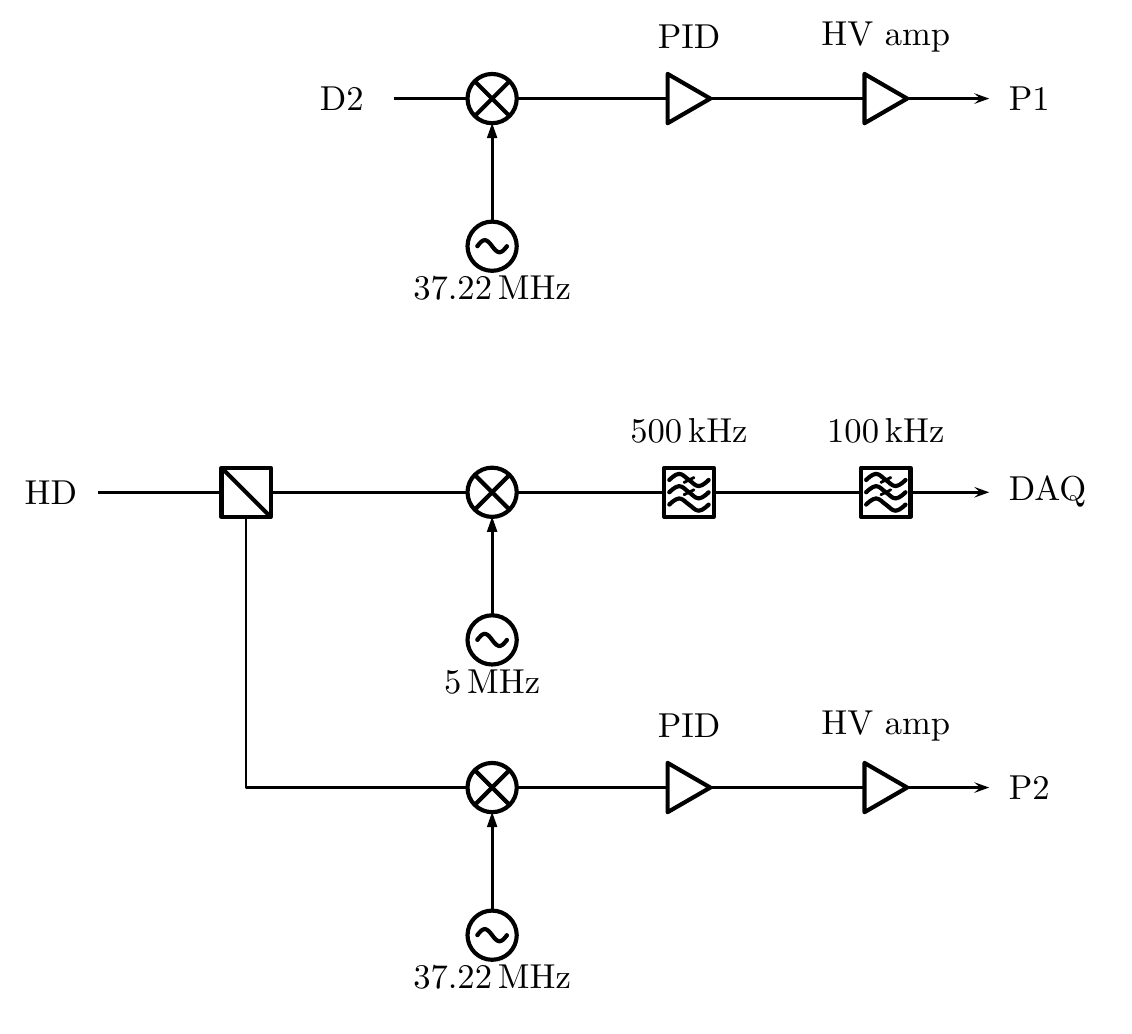}%
  \caption{%
  Diagram of the employed electronics.
  All shown devices were built in-house.
  External frequencies were generated by a digital synthesiser based on a \textsl{Analog Devices} AD9959 chip.
  The photocurrent from D2 was mixed with an electronic local oscillator at the same frequency as used for the phase modulation of the steering beam.
  Thereby we could reference the coherent state and the steering beam, carrying the squeezed state, to the beam providing the local oscillator.
  This holds as the steering beam and local oscillator were derived from the same source.
  The AC-coupled output of the homodyne photocurrent was split in two:
  The first half was mixed with a sinusoidal signal at \SI{5}{\mega\hertz} and low-pass filtered.
  This scheme mimics a spectrum analyser set to \SI{5}{\mega\hertz}.
  After down-mixing the signal, it was acquired by a 14-bit data acquisition (DAQ) card.
  To match the bandwidth of the card, a low-pass filter of \SI{100}{\kilo\hertz} was connected between the output of the homodyne detector and the DAQ card.
  The second half served for locking the homodyne detection on the amplitude quadrature.
  A stable measurement of the amplitude quadrature was ensured by a servomechanism.
  HV amp: High voltage amplifier.
  }
  \label{fig:srss_el_setup}
\end{figure}

\fig\ \ref{fig:srss_setup} illustrates a scaled version of the optical setup.
The electronic components involved to stabilise the measurement are summarised in \fig\ \ref{fig:srss_el_setup}.

It is convenient to separate the description into three parts:
The beam representing the squeezed state (drawn in red in \fig\ \ref{fig:srss_setup}), the one representing the coherent state (green) and finally the local oscillator (blue).

The local oscillator and the coherent state were derived from the infrared beam which was coupled into a polarisation maintaining single-mode fibre (\textsl{Thorlabs} P3-1064PM-FC-2).
A fibre-based beam splitter sent \SI{90}{\percent} of the power directly to a fibre collimator (\textsl{Thorlabs} TC12APC-1064).
To control the optical power and polarisation properties, photodetector D1 was placed after a removable mirror and a half wave plate / polarising beam splitter (HWP / PBS) combination.
The interference visibility between local oscillator and signal beam, which combined the squeezed- and the coherent state, was optimised by means of a telescope.
For the measurement run, the visibility read \SI{97}{\percent}, mainly limited by the large beam waist of \SI{3.4}{\milli\meter} of the signal beam due to a propagation length of circa \SI{7}{\meter} measured from the squeezer cavity \cite{Note2}.
The propagation length of \SI{7}{\meter} is due to the fact that the squeezing source and the actual experiment were built on two different optical tables.
Piezo P2 actuated a mirror to control the local oscillator's phase.
A HWP set the local oscillator power before combining it spatially with the signal beam.
The next HWP in combination with a PBS rotated both the local oscillator's and the signal beam's polarisation state, thereby splitting up the two beams into equal halves.
Finally, a homodyne detector converted the interference signal into a photocurrent.

Following the proposed input scheme (\fig\ 1 in the main text), the amplitude quadrature of the signal beam had to be detected.
Hence, the local oscillator and signal beam had to be locked at a fringe maximum or minimum.
For locking at this point, we implemented an electronic mixer to detected the interference at \SI{37.22}{\mega\hertz}, which was the modulation frequency of the steering beam's phase.
This technique is usually referred to as `AC lock' \cite{Freise2010}.

The smaller fraction of light split by the fibre-based PBS (\textsl{Thorlabs} PBC1064SM-APC) travelled through an electro-optical modulator (EOM) driven at \SI{5}{\mega\hertz} before coupled out by a fibre collimator.
The EOM (\textsl{Photline} NIR-MPX-LN-0.1-P-P-FA-FA) modulated the phase of the field, such that we prepared, at the given frequency, a coherent state in the phase quadrature.
This preparation scheme contrasts the concept illustrated in the main text, where an amplitude-displaced coherent state enters the interferometer.
In fact, as we prepared an \emph{amplitude} squeezed state, also the squeezed state does not correspond to the configuration in the main text.
Hence, the phase of both input states had to be rotated by \SI{90}{\degree} to match the illustrated configuration.
As outlined in the previous paragraph, a rotation of the relative phase can be achieved by implementing an AC lock.
Equal to the homodyne detection lock, the reference signal was provided by the phase modulation at \SI{37.22}{\mega\hertz} of the steering beam.
The interference visibility with the local oscillator measured \SI{99}{\percent} at the homodyne detector.
Depending on the amplitude of the applied modulation signal, the coherent state's average photon number was controlled.
To calibrate the voltage-to-photon-number conversion, a homodyne tomography was performed.

Next, we turn to the beam transferring the squeezed state, which we term ``squeezed beam''.
As mentioned above, the squeezer was operated at amplitude squeezing, i.e.\ de-amplification.
This operation is beneficial to the noise features of the steering beam, as an operation at amplification leads to an increase of technical noise from the laser source.
Furthermore, the decreased power helped to prevent saturation of the detector electronics.
In front of the squeezer cavity, the optical power of the steering beam measured \SI{400}{\micro\watt}.
The pump beam at \SI{532}{\nano\meter} had a power of \SI{45}{\milli\watt}.
To split a small fraction from the squeezed beam for interference with the coherent beam at detector D2, a HWP was placed in front of the PBS which combined the two beams spatially.
After this PBS, the squeezed beam was $s$ polarised, orthogonal to the $p$ polarisation of the coherent beam.
However, both beams share the same spatial mode, which was guaranteed by a visibility of \SI{99}{\percent} at detector D2.
In this situation, the two polarisation modes constitute the interferometer arms.
To delay one mode with respect to the other, i.e.\ to create the phase shift, a HWP was used.
Mounted in a remote-controlled rotation stage (\textsl{Thorlabs} K10CR1/M), we could control the experiment from a PC.
To mimic a common Mach--Zehnder interferometer, the relative phase shifts caused by reflections and the employed locking technique had to be considered.
Analysing the phase shifts by means of the Jones formalism \cite{Hecht2002} showed that an additional quarter wave plate was required to mimic the spatial Mach--Zehnder interferometer.
Its polarisation axis should be oriented parallel to either of the beams.

The important experimental parameters are summarised in table \ref{tab:srss_settings}.

\subsection{Data acquisition and processing}

The homodyne detector was a direct photocurrent subtraction design equipped with photodiodes with a quantum efficiency of $\eta_\text{qe} > \SI{99}{\percent}$.
The detector circuit featured three outputs: A DC output with a bandwidth of \SI{330}{\kilo\hertz}, an AC output with a high-pass filter of \SI{1}{\mega\hertz} and an output of a signal created by mixing the AC signal with an electronic local oscillator.
The latter signal was low-pass filtered at \SI{500}{\kilo\hertz} to provide a down-mixed signal for data acquisition.
In this way, the data was recorded on a PC without using an electronic spectrum analyser.
Since the coherent state was prepared by means of the EOM driven at a frequency of \SI{5}{\mega\hertz},
 the electronic local oscillator was set to the same frequency.
To minimise the required electrical power for the mixing process, the input for the electronic local oscillator was equipped with a resonant filter.

A digital oscilloscope with a PCIe interface (\textsl{GaGe} CSE8384) sampled the data from the homodyne detector.
It featured a bandwidth of \SI{100}{\kilo\hertz} at a sampling rate of \SI{500}{\kilo\hertz} and resolution of \SI{14}{bit}.
Per measurement point, \num[retain-unity-mantissa = false]{1e6} samples were acquired.
A measurement point was defined by the setting of the motorised HWP.
Centred about the null phase, \num{57} points were recorded every \SI{.9}{\degree} (\SI{15.7}{\milli\radian}).
Beyond this central region, \num{40} points were taken with an increment of \SI{2}{\degree}.
These points were used to validate the response of the stage actuating the HWP.
The overall procedure was repeated for seven different coherent state amplitudes.

The actual data processing was done as follows:
\begin{enumerate}
  \item Characterise the squeezed- and coherent state individually via homodyne tomography.
  Thus we account for all photons (potentially) interacting with the sample to faithfully calibrate the sensitivity \sens.
  The photons contributed by the local oscillator are not taken into account, as they do not enter the interferometer.
  \item Null the data for a correct zero phase.
  This step is important for further processing, as it symmetrises the data such that the \SI{0}{\degree} setting implies detecting a squeezed state.
  \item Choose a value for $a$ and sort the data of each measurement point according to
  \begin{equation}
    \label{eq:srss_dichotomy_operator_full}
    \ops \Pi = \lambda_0 \ops \Pi_0 + \lambda_1 \ops \Pi_1,
  \end{equation}
  with $\lambda_0 = 1 / \erf(\sqrt 2 a)$ and $\lambda_1 = 0$.
  As mentioned in the main text, $a$ was set to $1 / 2$.
  This step yields $\expct{\ops \Pi}$.
  \item Fit the function $\expct{\ops \Pi}$ to the experimental results.
  \item Calculate the variance at each data point.
  With an analytic expression of $\expct{\ops \Pi}$, the derivative in the sensitivity expression
  \begin{equation}
    \label{eq:sens}
    \sens = \sqrt{\var(\ops \Pi)} \bigg/ \abs*{\frac{\partial}{\partial\phi}\expct{\ops \Pi}},
  \end{equation}
  may be calculated.
  Given the variance and derivative at each data point, the sensitivity is computed.
  From the fit to $\expct{\ops \Pi}$ from the previous step, a comparison to an analytic expression can be derived.
  To justify the validity of this approach, \fig\ \ref{fig:sens_derivation} compares the sensitivity found by evaluating \eqref{eq:sens} as discussed, i.e.\ with a algebraic evaluation of the derivative, and an entirely numerical evaluation.
  \begin{figure}
    \centering
    \includegraphics{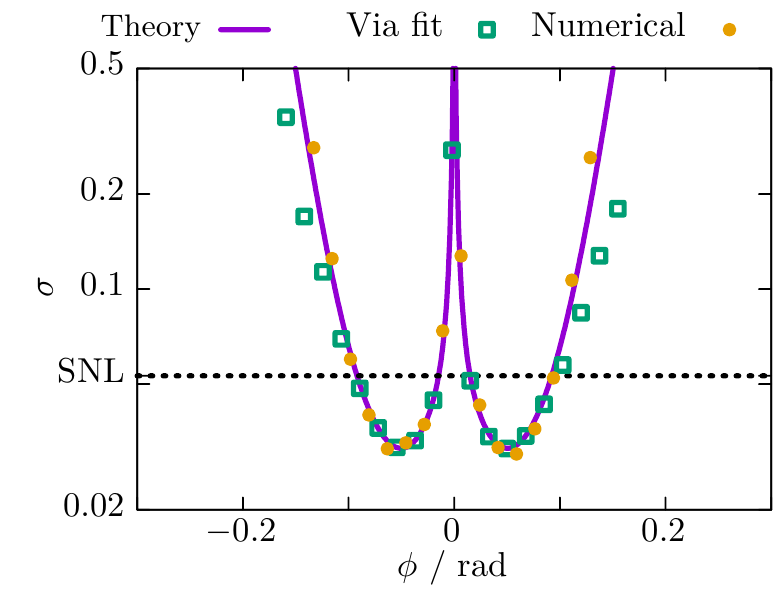} \hfill
    \includegraphics{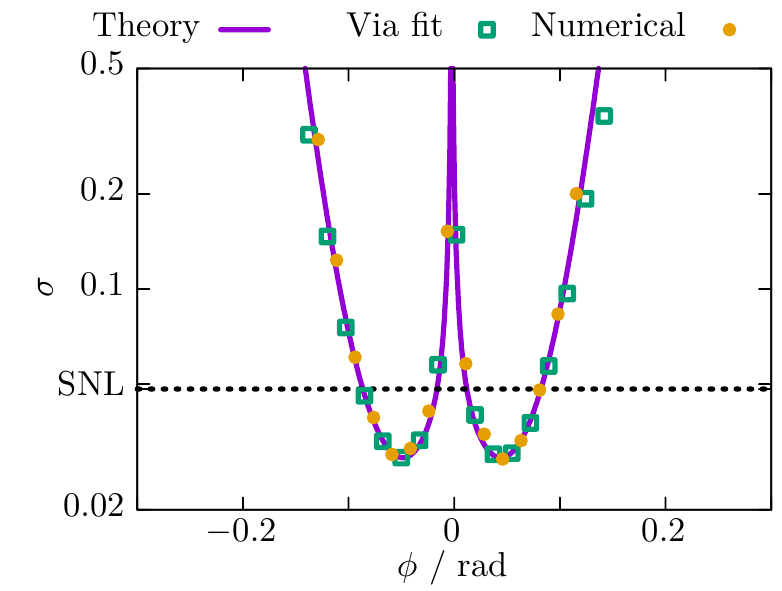}
    \caption{%
    Comparison of \sens\ where the derivative in \eqref{eq:sens} is evaluated by inserting the experimental data into a fit (square symbols) and where the derivative is evaluated numerically.
    The phase points of the ``all numerical'' approach are shifted, because the derivative is approximated at $(\phi_n + \phi_{n+1}) / 2$.
    On the other hand, using an algebraic expression for $\abs*{\frac{\partial}{\partial\phi}\expct{\ops \Pi}}$ allows for an evaluation at $\phi_n$.
    The left panel shows the sensitivity for $N \approx 355$, the right one shows it for $N \approx 427$ (\fig\ 4b in the main text).
    }
    \label{fig:sens_derivation}
  \end{figure}
\end{enumerate}

\renewcommand{\refname}{Supplementary References}

\end{document}